\documentclass[reprint,amsmath,amssymb,aps]{revtex4-1}

\usepackage{units}
\usepackage{ulem}
\usepackage{amsmath}
\usepackage{amssymb}
\usepackage{amsfonts}
\usepackage{epstopdf}
\usepackage{microtype}
\usepackage{graphicx} 			
\usepackage{bm}				
\usepackage{color}			
\usepackage{pstricks}			
\usepackage{float}			

    \newcommand{\be}		{\begin{equation}}	 	
    \newcommand{\ee}		{\end{equation}}	 	

    \newcommand{\Eq}[1]		{Eq.~(\ref{#1})}		
    \newcommand{\Eqs}[2]	{Eqs.~(\ref{#1}-\ref{#2})}	
    \newcommand{\eq}[1]		{(\ref{#1})}			
    \newcommand{\Fig}[1]	{Fig.~\ref{#1}}			

    \newcommand{\ex}[1]		{\mbox{e}^{#1}}			
    \newcommand{\re}[1]		{\mbox{Re}\left[#1\right]}	
    \newcommand{\pder}[3]	{\frac{\partial^{#1}#2}{\partial #3^{#1}}}		
    			
    \newcommand{\bx}		{\bm{x}}
    \newcommand{\br}		{\bm{r}}

    \newcommand{\gper}		{\gamma_\perp}			
    \newcommand{\gpar}		{\gamma_\parallel}		
    \newcommand{\muo}		{\mu_0}				

\newcommand{\eC}{\varepsilon}		
\newcommand{\pC}{\rho}			
\newcommand{\eCS}{a}			
\newcommand{\spc}{\br}			
\newcommand{\lasmodeS}{\Psi}		
\newcommand{\polmodeS}{p}		
\newcommand{\CFvec}{\varphi}		
\newcommand{\lasfreq}{\Omega}		
\newcommand{\atmfreq}{\lasfreq_a}	

\newcommand{\E}{\bm{E}}

\newcommand{\Ext}{E(\br,t)}
\newcommand{\Epxt}{E^+(\br,t)}

\newcommand{\Pxt}{P(\br,t)}

\newcommand{\Dxt}{D(\br,t)}

\begin{document}
\title{\textsf{Spectral method for efficient computation of time-dependent phenomena in complex lasers}}
\author{\textsf{O. Malik}}
\author{\textsf{K. G. Makris}}
\thanks{Current Address: \textsf{Institute for Theoretical Physics, Vienna University of Technology, Vienna, Austria.}}
\author{\textsf{H. E. T\"ureci}}
\affiliation{\textsf{Department of Electrical Engineering, Princeton University, Princeton, New Jersey 08544, USA}}
\date{\today}

\begin{abstract}
Studying time-dependent behavior in lasers is analytically difficult due to the saturating non-linearity inherent in the Maxwell-Bloch equations and numerically demanding because of the computational resources needed to discretize both time and space in conventional FDTD approaches. We describe here an efficient spectral method to overcome these shortcomings in complex lasers of arbitrary shape, gain medium distribution, and pumping profile. We apply this approach to a quasi-degenerate two-mode laser in different dynamical regimes and compare the results in the long-time limit to the Steady State Ab Initio Laser Theory (SALT), which is also built on a spectral method but makes a more specific ansatz about the long-time dynamical evolution of the semiclassical laser equations. Analyzing a parameter regime outside the known domain of validity of the stationary inversion approximation, we find that for only a narrow regime of pump powers the inversion is not stationary, and that this, as pump power is further increased, triggers a synchronization transition upon which the inversion becomes stationary again. We provide a detailed analysis of mode synchronization (aka cooperative frequency locking), revealing interesting dynamical features of such a laser system in the vicinity of the synchronization threshold. 
\end{abstract}
\maketitle

Lasers are very rich dynamical systems which exhibit various time-dependent phenomena characteristic of non-linear systems such as phase and mode locking, self-pulsing and breathing, and generally, spatio-temporal pattern formation and dynamical chaos. 
Almost all these effects can be understood and quantitatively studied using the semiclassical laser theory in the form of Maxwell-Bloch (MB) equations \cite{sargent_laser_1978,haken_light_1986,arecchi_laser_1972}, a set of coupled non-linear equations for the space- and time-dependent electric field amplitude $\Ext$, and the polarization and inversion of the gain medium $\Pxt$ and $\Dxt$. Early work made abundant use of spectral methods, where the field amplitudes entering the MB equations are expanded in a complete basis of spatial modes, reducing MB equations to a set of coupled non-linear ordinary differential equations for time-dependent amplitudes. These early theoretical investigations made a number of simplifying assumptions on the {\it spatial} aspects of the problem. The lasing modes were assumed to be simple (uniform, trigonometric, or gaussian) and unmodified from their passive cavity modes, and the openness (optical leakage) was taken into account phenomenologically. While these assumptions are sufficiently general to reproduce qualitatively almost all features of laser dynamics in macroscopic cavities, new laser systems have emerged in the past two decades that raised questions not easily addressable by these spectral approaches. 

Most novel laser systems are motivated by their deployment as compact and tunable light-sources for on-chip applications \cite{vahala_optical_2003}. Typically, these lasers feature complex sub-wavelength patterning of the cavity volume to employ light-confinement mechanisms that are based on optical interference (photonic band gap materials that may or may not include optical defects, random lasers) and/or total internal reflection (whispering gallery lasers, wave-chaotic lasers). Therefore these lasers feature spatially complex modes, typically in highly open geometries. In some cases, such as weakly scattering random lasers, it is not even clear where the boundary of the system is, and even what a mode means \cite{vanneste_lasing_2007}. In addition, many solid-state lasers are subject to spatially non-uniform pumping conditions and feature strong modal interactions \cite{chern_unidirectional_2003,kneissl_current-injection_2004,shakoor_room_2013,gmachl_high-power_1998,ge_enhancement_2014}. All these conditions can be modeled by appropriately setting up the original Maxwell-Bloch equations and solving the resulting non-linear partial differential equations (PDEs) in time-domain through various finite-difference-based numerical methods \cite{taflove_computational_2005,ziolkowski_ultrafast_1995,nagra_fdtd_1998}. A number of such powerful computational methods have been developed and employed to investigate the dynamics of complex laser systems, either solving the full set of MB equations \cite{jiang_time_2000,sebbah_random_2002,fratalocchi_mode_2008}, or the parabolic version obtained upon a slowly varying envelope (SVE) approximation in the time-domain, the so-called Schr\"odinger-Bloch (SB) equations \cite{harayama_theory_2005}. 

A more recent approach, the Steady-state Ab-initio Laser Theory (SALT) \cite{tureci_self-consistent_2006,tureci_strong_2008}, overcomes the rather expensive discretization of the spatial domain of a complex laser system in MB/SB-FDTD solvers by taking a spectral approach. The field amplitudes $\Ext$ are expressed in the Constant-Flux (CF) basis \cite{tureci_self-consistent_2006}, a set of non-Hermitian modes that {\it exactly} describe the steady-state field distribution in a {\it finite} and {\it open} domain under harmonic driving conditions \cite{claassen_constant_2010}. There are a number of advantages provided by this approach. (1) The steady-state multi-mode solution (to be defined precisely below) in the asymptotic infinite-time limit is obtained directly, without resorting to a time-domain simulation, (2) The exact solution of the MB equations in the {\it steady-state} is obtained through a modular two-stage procedure: in the first stage the {\it linear} problem corresponding to the determination of a CF basis is solved, and in the second stage this information is used to solve a set of algebraic transcendental equations \cite{claassen_constant_2010}. This allows the separation of spatial complexity (handled as a linear problem) from the computational non-linear problem, and perhaps more importantly obviates the need for the computational implementation of boundary conditions through various PML-variety  approaches \cite{berenger_perfectly_1994}. (3) SALT is also flexible enough to effectively account for spatially non-uniform pumping conditions \cite{tureci_strong_2008,ge_steady-state_2010,liertzer_pump-induced_2012,ge_enhancement_2014}, (4) can directly provide the farfield electric field distribution and spectrum \cite{ge_enhancement_2014}, and (5) over the past decade provided unique semi-analytic insight to fundamental problems in laser physics \cite{tureci_strong_2008,zaitsev_diagrammatic_2010,chong_general_2012,liertzer_pump-induced_2012,stano_suppression_2013,ge_enhancement_2014} that is harder to attain through brute-force computational approaches. 

Despite the success of SALT in the treatment of complex laser systems, it has certain well-known limitations. The key assumption of the theory is the stationarity of the inversion $D(\br,t)$  \cite{ge_quantitative_2008} (or more generally, the level populations \cite{cerjan_steady-state_2015}). The inversion is however never truly stationary, but in a certain regime of parameters the non-stationary corrections are systematically very small and can be neglected. To be more specific, the non-stationary corrections \cite{ge_quantitative_2008}, as discussed in detail below, are order $\gpar/\Delta$ where $\Delta$ is the smallest frequency difference of the lasing modes (typically slightly different from the free spectral range of the cold cavity) and $\gpar$ is the inversion relaxation rate. However, $\Delta$ depends on the pump strength, and at larger powers can become smaller than $\gpar$ due to non-linear effects. As a consequence, the corrections to SALT are not going to be small, and can lead to {\it qualitatively} different behavior. As shown here, such a scenario can take place under unusual circumstances where the spectrum of the cavity contains quasi-doublets (typically protected through a discrete spatial symmetry of the cavity) that are spectrally spaced apart at a distance ($\Delta_2$) that is larger than the splitting of the doublets ($\Delta$), as shown in \Fig{fig:cav_char}. Under such circumstances the lasing modes of the doublet-pair most favored by the gain curve can lock to each other and synchronize as pump power is increased through an effect called 'cooperative frequency locking' \cite{lugiato_cooperative_1988}. Yet even then, as we will show, the Stationary Inversion Approximation (SIA) fails  only in a very limited pump power range near the synchronization threshold, and is valid for most of the pump power range below and above this threshold. 

Thus it is of interest not only to understand the validity of the SIA  under various circumstances, but also to develop a spectral method that is in principle not limited by any approximations such as the SIA. Such a technique should be able to capture any intrinsically dynamical behavior of complex lasers. We present such a technique in this article, and discuss precisely how SIA and thus SALT may fail in certain limited parameter regimes. 

Just as SALT, the new Constant Flux Time Domain (CFTD) technique presented here provides a versatile tool for calculating lasing thresholds, spectra, and modal distributions in the multi-mode regime for complex lasers including random \cite{tureci_strong_2008}, semiconductor \cite{redding_low_2015}, photonic crystal surface-emitting \cite{chua_low-threshold_2011}, and photonic molecule lasers \cite{brandstetter_reversing_2014}. Unlike SALT, it can capture transient regimes, locking and synchronization, various dynamical instabilities \cite{andreasen_coherent_2011}, as well as dynamical chaos and generally, spatio-temporal pattern formation. 

In Section I we provide an overview of our theoretical approach, outline key approximations, and establish the CFTD-SALT correspondence. In Section II, we provide a comparative study of CFTD and SALT for a two-mode quasi-degenerate laser in two different regimes of parameters. Keeping all other parameters the same, we analyze the steady-state dynamics of this laser for small $\gpar = 0.001$ ($\gpar/\Delta = 0.0038$) where the SIA is valid, and then for $\gpar = 1$ ($\gpar/\Delta = 3.8037$) where SIA can not be guaranteed. Indeed, in the latter regime we illustrate that the inversion is non-stationary for a narrow range of pump powers, and show how this destabilizes the stationary emission and ultimately triggers the synchronization of the two modes, to return to a dynamical regime where inversion is again stationary. 

\section{Non-hermitian spectral approach to laser dynamics}

We start with the following form of the Maxwell-Bloch equations \cite{haken_light_1986} for the scalar electric field amplitude $\Ext$, polarization $\Pxt$ and inversion density $\Dxt$:
\begin{gather}
	\nabla^2 E^+ - \frac{n^2}{c^2}\ddot{E}^+ = \muo\ddot{P}^+, 			\label{eqn:MEe} \\
	\dot{P}^+ = -(i\atmfreq + \gper)P^+ - i\frac{g^2}{\hbar}E^+D, 			\label{eqn:MEp} \\
	\dot{D}   = \gpar[D_0(\bx) - D] + i\frac{2}{\hbar}[E^+(P^+)^* - (E^+)^*P^+].	\label{eqn:MEd}
\end{gather}
Here $E=E^+ + E^-$, $P = P^+ + P^-$ and we used the rotating wave approximation (RWA), valid when the frequencies of the aforementioned fields ($\sim \atmfreq$) are much larger than their relaxation rates (controlled by $\gpar$ in Class A and B lasers \cite{ge_gain-tunable_2013}), typically well-satisfied in the optical regime. The laser cavity is characterized by the complex-valued refractive index distribution $n(\br)$. We have in mind a quasi-2D geometry in which case the scalar field $E(\br,t)$ denotes the $z$-component of the electric field for transverse magnetic (TM) polarization, and $n(\br)$ represents the effective index \cite{lebental_inferring_2007}. In the inversion equation, $D_0(\br)$ represents the possibly spatially inhomogeneous pump distribution. We note that the description in \Eqs{eqn:MEp}{eqn:MEd} is sufficiently general to describe the salient features of various gain media characterized by a single dominant optical transition frequency, including quantum cascade-based lasers (see Supplementary Information in \cite{ge_enhancement_2014}). The remaining parameters are as follows: $\gper$ and $\gpar$ are the polarization and inversion decay rates, $\atmfreq$ is the center frequency of the gain curve, $g$ is the dipole moment of the individual two-level emitters forming the gain medium, $\muo$ is the magnetic permeability, and $c$ is the speed of light.

In the standard spectral approach \cite{haken_laser_1984_ch5}, the electric field and polarization are expanded in a complete set of states $\phi_m(\br)$ e.g. $\Epxt = \sum_m c_m(t) \phi_m(\br)$, with $\phi_m$ satisfying $\nabla^2 \phi_m (\br) = -n^2(\br) \, (\omega_m^2/c^2) \, \phi_m (\br)$ with a boundary condition at the cavity walls $\partial D$ that gives rise to a Hermitian boundary value problem, and hence a complete set of orthogonal states $\{ \phi_m \}$ with real-valued frequencies $\{ \omega_m \}$. There are two crucial shortcomings of this approach. The first is that a phenomenological decay rate has to be added by hand to the equations \cite{haken_laser_1984_loss} in order for a well-defined steady-state to exist. As an additional consequence, there is no systematic way to extend the solution to the exterior of the cavity, where the fields are actually measured. A second shortcoming with this approach is that spatial hole burning interactions can only be captured perturbatively in the electric field amplitude, or else through an adiabatic elimination of the gain medium degrees of freedom.

\begin{figure}[h]
\includegraphics[clip,width=1.0\linewidth]{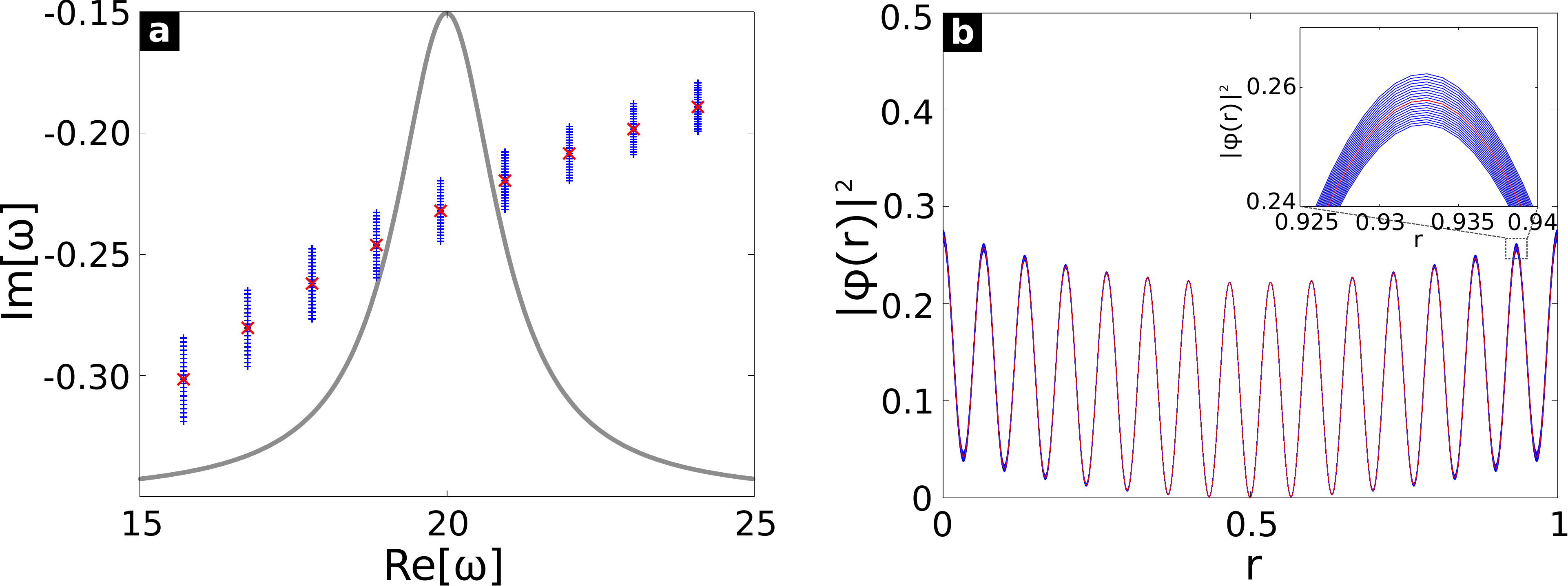}
\caption{(a) CF Eigenvalues and their variation as a function of the external frequency are shown for a 1D cavity with $n=3$, $\atmfreq = 20$, $\gper = 1$, and $n\atmfreq L/c = 60$ (see description of parameters in text). Closely spaced blue '+' markers indicate the variation in eigenvalues within the range $\Omega = \atmfreq \pm \gper$ and the red 'x' marker indicates the eigenvalue calculated at $\Omega = \atmfreq$. The shift in the real part of these eigenvalues is negligibly small. (b) The variation in the CF eigenstate associated with the smallest eigenvalue shown in (a). Blue lines show the eigenstate calculated in the same range as (a) while the red line shows the eigenstate calculated at $\Omega = \atmfreq$. The inset zooms in on the peak enclosed in the dotted rectangle, closely showing the very small variation in the eigenstates as a function of $\Omega$. }
\label{fig:eigenspread}
\end{figure}

Here, we extend this spectral approach to a consistent mathematical framework, by first expanding the electric and polarization fields in terms of CF states \cite{tureci_self-consistent_2006} through the following ansatz: 
\begin{align}
	E^+(\spc,t) &= \sum_n \eC_n(t)\CFvec_n(\spc,\atmfreq)\ex{-i\atmfreq t}, \label{eq:Eans} \\
	P^+(\spc,t) &= \sum_n \pC_n(t)\CFvec_n(\spc,\atmfreq)\ex{-i\atmfreq t}  \label{eq:Pans}.
\end{align}
The biorthogonal set of CF states $\{\CFvec_m (\br,\Omega)\}$ is the solution to the Laplace eigenvalue problem $\nabla^2 \CFvec_m = -n^2(\br)  \, (\omega_m^2/c^2) \, \CFvec_m$ with outgoing boundary conditions $\pder{}{\CFvec_m}{r} \sim i (\Omega/c) \CFvec_m$ as $r \rightarrow \infty$. The set of CF states is the exact non-Hermitian  basis to expand the fields in an arbitrary open geometry described by $n(\br)$ that is excited by an arbitrary spatial distribution of monochromatic sources at frequency $\Omega$ \cite{claassen_constant_2010}. The solution to this boundary value problem leads to a complex-valued spectrum ${\omega_m}$ and associated eigenmodes $\CFvec_m$ that parametrically depend on the excitation frequency $\Omega$ (see \Fig{fig:eigenspread} for an example of this parametric dependence). The imaginary part of ${\omega_m}$ provides the crucial mode-dependent losses, either through optical leakage out of $\partial D$ or the material absorption described by the imaginary part of $n(\br)$.

A crucial computationally important detail here is that the computational domain of the CF problem can be reduced to a "last scattering surface'' $\partial D$ that can be chosen to be the minimal volume that includes all the relevant scattering elements. In practice \cite{tureci_strong_2008}, $\partial D$ is chosen to be the minimal circular boundary (in 2D) that includes all the spatial inhomogeneities of $n(\br)$. Therefore, by construction the relevant open boundary conditions are {\it exactly} satisfied through the use of the CF basis in the expansion  \Eq{eq:Eans}. In addition, CF states can be analytically continued straightforwardly outside $\partial D$ and hence the fields, and in particular the electromagnetic flux and the measured spectrum, can be calculated exactly in the farfield \cite{ge_enhancement_2014}. 

In the ansatz \eq{eq:Eans}-\eq{eq:Pans} the time-dependence of each field variable is entirely encapsulated in its respective coefficients $\eC_n$ and $\pC_n$. For computational efficiency, we factor out the fast oscillation at atomic frequency $\atmfreq$. The spatial dependence is entirely captured by the CF states, which are calculated, in a departure from previous applications of the CF basis, only at $\atmfreq$. This is a very good and well-controlled approximation, for the CF states and frequencies $\{\CFvec_m (\br,\Omega) , \omega_m(\Omega) \}$ typically change slowly when the excitation frequency $\Omega$ is varied, see for an example \Fig{fig:eigenspread}. This is in fact one of the  crucial factors in the computational efficiency of SALT \cite{tureci_strong_2008}. We note that it is possible to choose an unusual geometry where for a certain narrow regime of parameters ($\Omega$) this assumption may fail, but generally this should be taken as a hint that some extraordinary spatial physics is present in the system that may give rise e.g. to an exceptional point \cite{liertzer_pump-induced_2012}. 

With the above ansatz of \Eqs{eq:Eans}{eq:Pans} inserted into \Eqs{eqn:MEe}{eqn:MEd}, we can derive the following equations of motion for the {\it time-dependent dimensionless} coefficients $\tilde{\eC}(t)$, $\tilde{\pC}(t)$, $\tilde{D}_{mn}(t)$:
\begin{align}
	\dot{\tilde{\eC}}_m  ={}& \frac{i}{2\atmfreq}(\atmfreq^2 - \omega_m^2)\tilde{\eC}_m + i\frac{\atmfreq}{2}\sum_n B_{mn}\,\tilde{\pC}_n, 	\label{eq:ETD} \\
	\dot{\tilde{\pC}}_m  ={}& -\gper\tilde{\pC}_m - i\gper\sum_n \tilde{D}_{mn} \, \tilde{\eC}_n, 							\label{eq:PTD} \\
	\begin{split} 
		\dot{\tilde{D}}_{mn} ={}& \gpar(\tilde{D}_{0,mn} - \tilde{D}_{mn}) \\
			        &+ i\frac{\gpar}{2}(\sum_{rs} A_{mrsn} \tilde{\eC}_r\tilde{\pC}_s^* - A'_{mrsn} \tilde{\eC}_r^*\tilde{\pC}_s ).	\label{eq:DTD}
	\end{split}
\end{align}
Here we introduced the inversion matrix $D_{mn} (t) = \int_{\text{cavity}}~d\spc~n(\spc)\CFvec_m(\spc) D (\spc,t)\CFvec_n(\spc)$, a set of space-independent coefficients describing the mode-projected inversion distribution. While the inversion $D(\spc,t)$ itself is real-valued, the coefficients $D_{mn}$ are in general complex-valued. All the variables are rendered dimensionless through $\tilde{\eC}_m = \eC_m/E_c$, $\tilde{\pC}_m = \pC_m/P_c$, and $\tilde{D}_{mn} = D_{mn}/D_c$ using the following scale factors that contain all the units:
\begin{equation}
	E_c = \frac{\hbar\sqrt{\gper\gpar L}}{2g},~~P_c = \frac{E_c}{\muo c^2},~~ D_c = \frac{\hbar\gper}{\muo g^2c^2}.
	\label{eq:scalingfacts}
\end{equation}
Furthermore, time and decay rates are scaled by the {\it effective} cavity round-trip time $t_{RT} = nL/c$ ($n$ can be taken to be the spatially averaged effective index, and $L=V^{1/3}$ for a cavity with volume $V$). The key step in obtaining \Eqs{eq:ETD}{eq:DTD} is the elimination of the spatial dependence of each field vector by utilizing the biorthogonality of the CF basis vectors. We will drop the tildes henceforth.
\be
\int_{\text{cavity}}~d\spc~n(\spc) \CFvec_m(\spc) \CFvec_n(\spc) = \delta_{mn}
\ee
In contrast to a Hermitian orthogonality relation, this inner product does not contain a complex conjugation. This is a consequence of the dual modes (left eigenvectors) $\{ \bar{\CFvec}_m \}$ satisfying the relationship $\bar{\CFvec}_m = \CFvec_m^*$ \cite{tureci_self-consistent_2006,claassen_constant_2010}. 
 
This step produces the following unitless complex-valued parameters appearing in the above equations,
\begin{align}
	A_{mrsn}  &= L\int_{\text{cavity}}~d\spc~n(\spc)\CFvec_m(\spc)\CFvec_r(\spc)\CFvec_s^*(\spc)\CFvec_n(\spc), 	\label{eq:Amrsn}  \\
	A'_{mrsn} &= L\int_{\text{cavity}}~d\spc~n(\spc)\CFvec_m(\spc)\CFvec_r^*(\spc)\CFvec_s(\spc)\CFvec_n(\spc),	\label{eq:Apmrsn} \\
	B_{mn}	  &= \int_{\text{cavity}}~d\spc~\CFvec_m(\spc)\CFvec_n(\spc),						\label{eq:Bmn}	  \\
	D_{0,mn}  &= \int_{\text{cavity}}~d\spc~n(\spc)\CFvec_m(\spc) D_0(\spc)\CFvec_n(\spc).				\label{eq:Domn}
\end{align}
Here, $A_{mrsn}$ and $A'_{mrsn}$ can be seen as a generalization of the inverse mode volume in the Hermitian version of the single-mode laser problem. Interestingly,  $B_{mn}$ is not diagonal unless the index is uniform across the cavity. The effective mode-projected pump parameter is given by $\tilde{D}_{0,mn}$ and is the most critical parameter here. These overlap integrals \Eqs{eq:Amrsn}{eq:Domn} are calculated prior to numerically solving the time-dependent system of coupled equations in \Eqs{eq:ETD}{eq:DTD} and they encapsulate the impact of the resonator modal structure on modal interactions. 

An important aspect of the above spectral formulation of semiclassical laser equations is that it takes into account modal interactions through spatial hole burning exactly. Majority of the past spectral methods (with the exception of SALT), account for interactions only perturbatively and generally to third order in the electric field amplitude. This approximation, as pointed out first in \cite{tureci_mode_2005} and later in quantitative detail discussed in \cite{ge_quantitative_2008}, is only valid near the lowest laser threshold, and generally severely underestimates the number of lasing modes at higher pump powers. 

As long as the parametric variations of the CF basis is small within a window $\gper$ of $\atmfreq$, the \Eqs{eq:ETD}{eq:DTD} are exact up to the slowly varying envelope approximation used in \Eq{eq:ETD} to remove second order time derivatives in $\tilde{\eC}_n$ and $\tilde{\pC}_n$. The impact of the latter in SALT has been quantified perviously \cite{ge_quantitative_2008} and was shown to introduce small inaccuracies in the calculation of steady-state lasing characteristics, but was not found to lead to any qualitative differences. This approximation is not critical to the success of the method as discussed below, and can easily be undone at the expense of introducing additional fields.

In the next section, our goal is twofold. We would first like to benchmark SALT against the CF-projected time-dependent laser equations \Eqs{eq:ETD}{eq:DTD} (CFTD) in the regime of parameters where SALT is known to be accurate. Next, we investigate a regime accessed by the change of a single parameter, $\gpar$, leaving all other parameters the same, where the SIA is suspect. Here we encounter a narrow regime of pump powers where the system is critical and unstable towards a synchronized oscillation regime. In this regime that, for the special cavity configuration of \Fig{fig:cav_char}, occurs at extremely high pump power (about 25 times the lowest threshold), SALT fails to capture the underlying dynamics {\it qualitatively}. Interestingly, below and above this narrow regime of pump powers, the SIA is valid and SALT is accurate. 

A second aim of the following discussion is to present an accurate picture of the synchronization transition, known as cooperative frequency locking \cite{lugiato_cooperative_1988}. Our theoretical result is able to accurately capture the interesting dynamical regime around the critical pump power for locking, experimentally observed for the first time in 1988 \cite{tamm_frequency_1988}.

\section{Benchmarking SALT against CFTD: the two-mode quasi-degenerate laser}

\begin{figure}[h]
\includegraphics[clip,width=1.0\linewidth]{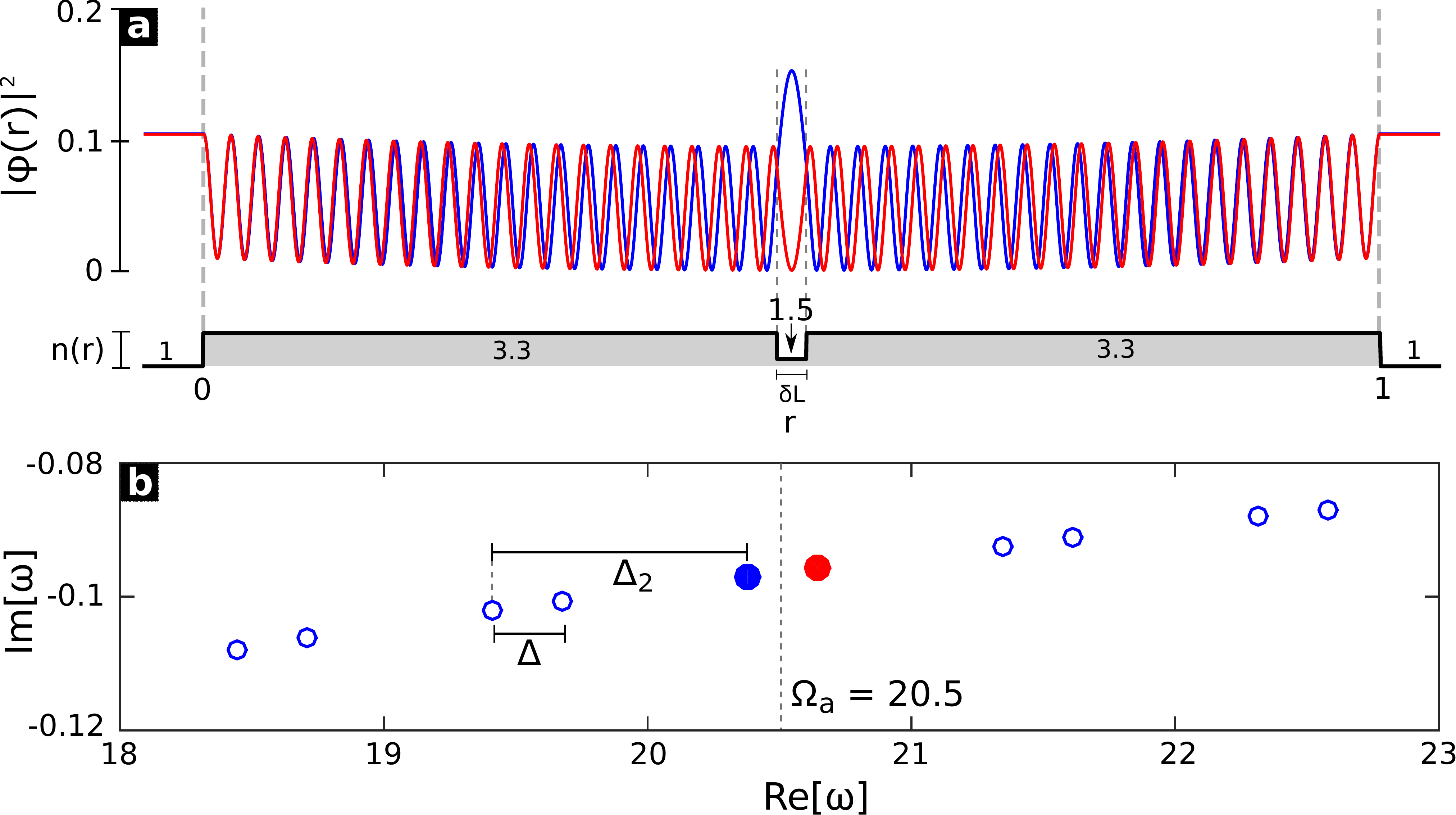}
\caption{(a) The refractive index distribution of the cavity described in the text is shown as a function of space ($r$), scaled by cavity length $L$. The CF eigenvectors color-matched with the CF eigenvalues marked in (b) are plotted inside and outside the cavity; dashed gray lines on each end mark the cavity faces. Note that the CF mode with an anti-node in the low refractive index region shows the expected large magnitude due to reduced index. (b) The 10 eigenvalues closest to gain center are shown for the cavity shown in (a); filled red/blue circles mark the two eigenvalues used in the calculations in this section: $\omega_1 = 20.3764 - 0.0970i$ and $\omega_2 = 20.6393 - 0.0957i$. The spacing between them is $\Delta = \re{\omega_2 - \omega_1} = 0.2629$ and their spacing from the adjacent eigenvalue pairs is $\Delta_2 \approx 0.7$. }
\label{fig:cav_char}
\end{figure}

In this section, laser dynamics is investigated for a quasi-degenerate 1D cavity. It consists of a dielectric slab with refractive index $n=3.3$ which sandwiches symmetrically a layer of index $n=1.5$ of thickness $\delta L$ (see \Fig{fig:cav_char}). The gain curve is centered at $\atmfreq = 20.5$, $\gper = 8$, and $n\atmfreq L/c \approx 135$. This particular choice of $\gper$ ensures a flat gain experienced by both of the cavity resonances included in the calculations below, significantly reducing the effect of gain-pulling in the time-dynamical scenario where lasing mode frequencies shift strongly with the pump power.

Choosing $\delta L/L = 1/40$, the resonances of the cavity come in quasi-doublets which are separated from each other by a relatively large spectral range (See \Fig{fig:cav_char}). These conditions are ideal to consider two regimes, one in which the SIA is valid (Regime A) and another where it cannot be guaranteed (Regime B), by changing the value of a single parameter, $\gpar$, and leaving all other parameters identical. 

In regime A ($\gpar \ll \Delta \ll \gper$), the assumptions underlying SIA are valid and SALT and CFTD results should agree quantitatively \cite{ge_quantitative_2008}. We will first set up the correspondence between SALT and CFTD variables in the steady-state and then demonstrate excellent agreement between the two methods using the 2-mode quasi-degenerate laser as an example.

In regime B ($\Delta \ll \gpar \ll \gper$) however, accessed here by changing $\gpar$, the stationary inversion approximation can not be guaranteed. Indeed, while at low powers the laser oscillates in two frequencies (``two-mode lasing"), above a critical pump power $D_0 = D_\text{sync}^\text{th}$ corresponding to the {\it threshold for synchronization}, these two frequencies lock and a single frequency remains. Just prior to synchronization, a close up at the power spectrum of various dynamical variables of CFTD in \Eqs{eq:ETD}{eq:DTD} reveals that close to the synchronization threshold the SIA breaks down. The SIA remains valid generally however, breaking down in a very interesting way but only within a narrow range of pump powers.

\subsection*{SALT-CFTD correspondence}

The strength of a steady-state approach like SALT \cite{tureci_self-consistent_2006} is that it directly delivers the frequencies as well as the intra- and extra-cavity field amplitudes as functions of the pump power $D_0$. As such, it is not immediately clear how the SALT variables are related to the CFTD variables $\eC_m$, $\pC_m$ and $D_{mn}$. In this subsection, we will set up this correspondence when this correspondence exists, and then compare SALT and CFTD results in the following subsection.

SALT is obtained by making a more specific ansatz for the long-time solution of MB equations than that for CFTD:
\begin{align}
	E^+(\spc,t) &= \sum_{\mu} \lasmodeS^{(\mu)}(\spc)\ex{-i\lasfreq^{(\mu)} t}, \label{eq:EansS} \\
	P^+(\spc,t) &= \sum_{\mu} \polmodeS^{(\mu)}(\spc)\ex{-i\lasfreq^{(\mu)} t}.  \label{eq:PansS}
\end{align}
The crucial point here is the assumption of a specific form for the exact time-dependence once steady-state is reached (compare \Eq{eq:EansS} to \Eq{eq:Eans}). The fields are assumed to be expandable in a discrete Fourier representation with a finite number of laser frequencies $\lasfreq^{(\mu)}$, which are unknown and to be determined. $\lasmodeS^{(\mu)}(\spc)$ are the spatial field amplitudes corresponding to the exact (non-linear) lasing modes, also to be determined through the SALT equations:
\begin{align}
\left[ \nabla^2 + (\epsilon_c(\vec{r}) + \epsilon_g(\vec{r}))\Omega_\mu^2 \right]\Psi_\mu(\vec{r}) = 0, ~ ~ \vec{r} \in \text{cavity} \label{eq:SALT} \\
\epsilon_g(\vec{r}) = \frac{\gper}{\Omega_\mu - \omega_a + i\gper}\frac{D_0 (\vec{r})}{1+\sum_{\nu=1}^N\Gamma_\nu|\Psi_\nu(\vec{r})|^2}. \label{eq:epsg}
\end{align}
Here $\Gamma_\nu = \gper^2/[\gper^2 + (\Omega_\mu - \atmfreq)^2]$. Note that the polarization spatial amplitudes $\polmodeS^{(\mu)}(\spc)$ can be directly related to $\lasmodeS^{(\mu)}(\spc)$ and do not show up in the final set of equations to be solved. Also note that in contrast to CFTD, the index $\mu$ specifically identifies lasing modes oscillating at distinct frequencies $\lasfreq^{(\mu)}$ (as opposed to spatial modes). The {\it time-independent} SALT equations \Eq{eq:SALT} are then solved by projecting each lasing mode $\lasmodeS^{(\mu)}(\spc)$ into a set of CF states for the associated frequency of oscillation $\lasfreq^{(\mu)}$:
\begin{equation}
	\lasmodeS^{(\mu)}(\spc) = \sum_n \eCS_n^{(\mu)}\CFvec_n(\spc,\lasfreq^{(\mu)}).
\end{equation}
The SALT-CFTD correspondence is unveiled by assuming $\CFvec_n(\spc,\lasfreq^{(\mu)}) \approx \CFvec_n(\spc, \atmfreq)$, which as  discussed before, is generally a good approximation. In that case, 
\begin{equation}
\eC_n (t) = \ex{ i \atmfreq t} \sum_\mu \eCS_n^{(\mu)} \ex{-i\lasfreq^{(\mu)} t}
\end{equation}
where the coefficients on the left and right hand side are the CFTD and SALT variables, respectively. 

Additional insight is obtained by asking what assumptions SALT makes about the solution of CFTD equations \Eqs{eq:ETD}{eq:DTD}, that are more general. SALT corresponds to specific long-time solutions of the CFTD equations for which $\eC_n (t) = \sum_\mu \eC^\mu_n \ex{-i\Omega_\mu t}$,  $\pC_n (t) = \sum_\mu \pC^\mu_n \ex{-i\Omega_\mu t}$ and $\dot{D}_{mn} = 0$. The last assumption is the mode-projected version of the SIA and one of the consequences is that $\gpar$ drops out of the equations. That doesn't however mean that SALT solutions do not depend on $\gpar$, but rather that the entire $\gpar$-dependence of SALT solutions is contained in the particular scaling of the electric field \Eq{eq:scalingfacts}. Of course, being exactly equivalent to MBE equations up to the aforementioned approximations, the CFTD equations permit far more general solutions, one of which we will encounter further below.


For an $N$-mode CFTD calculation where $N$ is the number of modes, a CF basis of equivalent dimension must be constructed, and $N^2+2N$ equations must be solved. For a 2-mode calculation, this amounts to 2 equations each for the electric and polarization fields, and a total of 4 equations for the diagonal and off-diagonal elements of inversion. Below we will discuss the two-mode regime for the quasi-degenerate 1D laser we introduced before (\Fig{fig:cav_char}).

\subsection*{Regime A: Stationary inversion}

We use the parameters quoted at the beginning of Section II and take $\gpar = 0.001$. Right at the onset of the second mode this gives $\gpar/\Delta = 0.0038$. $\Delta$ only slightly changes in the calculated interval of pump powers (see \Fig{fig:2mode}(b)) and the assumptions underlying the SIA remain rigorously valid throughout.  In the figures below, we use a normalized pump $\beta = D_0/D_{th}$ where $D_{th}$ is the lasing threshold.

\begin{figure}[h]
\includegraphics[clip,width=1.0\linewidth]{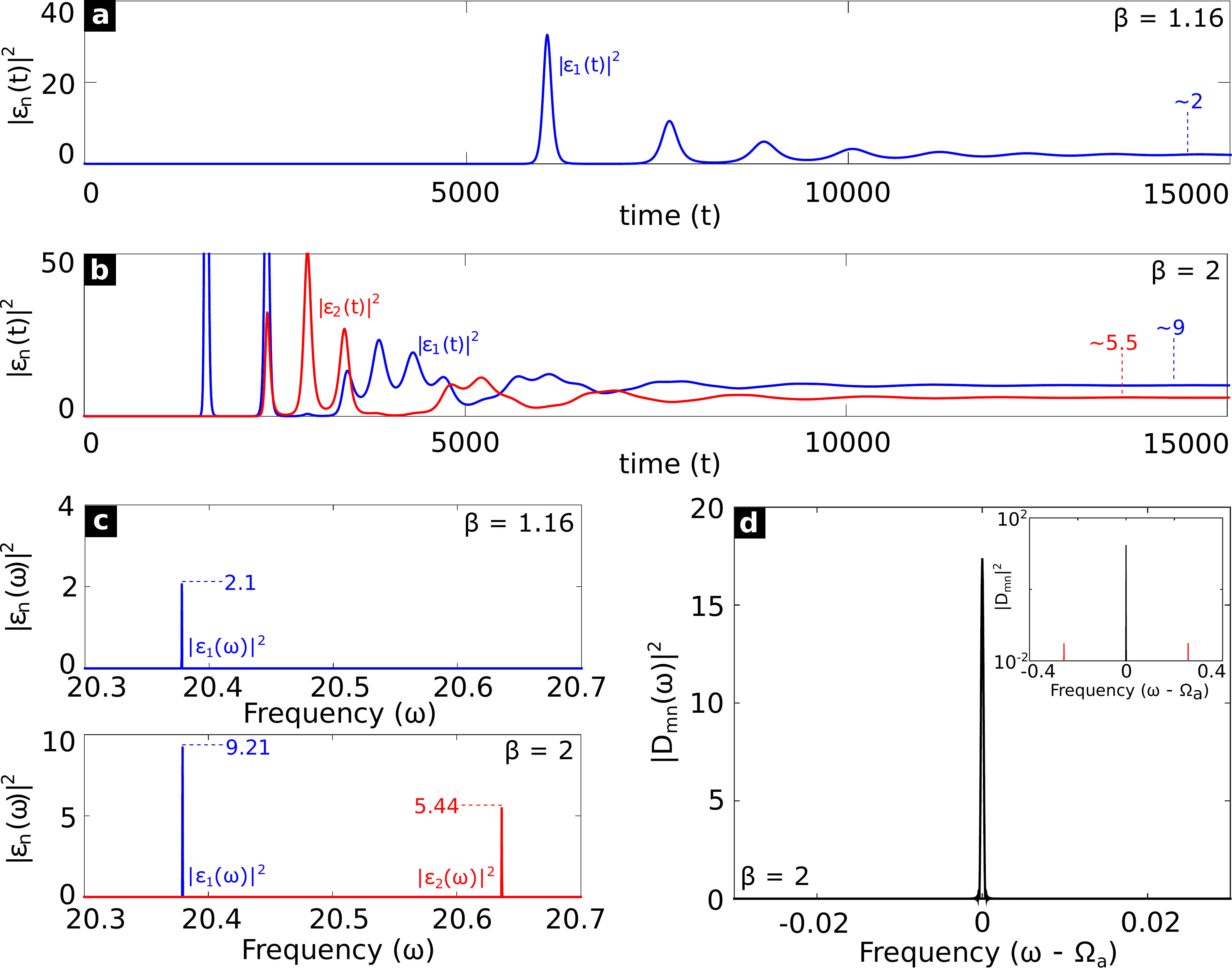}
\caption{(a) The time-domain behavior of $|\eC_1(t)|^2$ (blue) at $\beta=1.16$ in the single-mode lasing regime. (b) The time-domain behavior of $|\eC_1(t)|^2$ (blue) and $|\eC_2(t)|^2$ (red) in the multi-mode regime. Their approximate steady-state values are labeled. (c) The PSDs of $\eC_1(t)$ (blue) and $\eC_2(t)$ (red) at the same $\beta$ values as above; the labeled values compare well to those marked in (a) and (b) and in \Fig{fig:2mode}(a). (d) The spectral content of the diagonal inversion element $D_{11}$ (black peak); inset log-plot also plots the off-diagonal element $D_{12}$ (smaller red side-peaks) and shows that it's nearly 3 orders of magnitude smaller than $D_{11}$.}
\label{fig:CFTD}
\end{figure}

CFTD calculations show that both $|\eC_1(t)|^2$ and $|\eC_2(t)|^2$ reach steady state after initial transients die out. Some sample time-series are shown in \Fig{fig:CFTD}(a,b) for two different pump powers $\beta=1.16$ and $\beta=2$. These findings indicate that there is a single frequency in both $\eC_1(t)$ and $\eC_2(t)$, as confirmed in the respective power spectra shown in \Fig{fig:CFTD}(c). Shown in \Fig{fig:CFTD}(d), $|D_{mn}(\omega)|^2$ is the power spectral density (PSD) of $D_{mn}(t)$ (itself not shown). Here we plot $|D_{11}(\omega)|^2$ and $|D_{12}(\omega)|^2$ only. Further detail is shown in the inset which zooms out and shows in logarithmic scale that the non-stationary components in $D_{12}$ (red peaks) are suppressed by more than three orders of magnitude with respect to the static component of $D_{11}$ (black peak). The smallness of these side-peaks indicates that the SIA is an excellent approximation and the SALT-CFTD correspondence should be possible, which is what we do next. 

The SALT calculation containing two lasing modes expanded into a basis of two CF eigenvectors will contain four coefficients ($\eCS_1^{(1)}$, $\eCS_2^{(1)}$, $\eCS_1^{(2)}$, $\eCS_2^{(2)}$) and two lasing frequencies ($\lasfreq^{(1)}$, $\lasfreq^{(2)}$). As discussed in the previous section, in the steady-state, the information contained in these SALT variables can be retrieved from the two time-dependent CFTD variables ($\eC_1$, $\eC_2$). To do so, we simply expand and rearrange the SALT ansatz for two lasing modes,
\begin{align}
	\E^+(\spc,t) 	&= (\eCS_1^{(1)}\ex{-i\lasfreq^{(1)} t} + \eCS_1^{(2)}\ex{-i\lasfreq^{(2)} t})\CFvec_1(\spc,\atmfreq) \\
		      	&+ (\eCS_2^{(1)}\ex{-i\lasfreq^{(1)} t} + \eCS_2^{(2)}\ex{-i\lasfreq^{(2)} t})\CFvec_2(\spc,\atmfreq) \\ &= \eC_1(t)\CFvec_1(\spc,\atmfreq) + \eC_2(t)\CFvec_2(\spc,\atmfreq)
\end{align}
The CFTD results imply that $\eC_n (t) \approx  \eCS_n^{(n)} \ex{-i\lasfreq^{(n)} t}$, in other words $\eCS_n^{(m)} \approx 0$ for $n \neq m$. In SALT language, this means that the single-pole approximation is valid throughout the calculated regime -- the CF eigenvectors calculated for a cold cavity very closely represent the two lasing modes $ \lasmodeS^{(\mu=1,2)}(\spc)$, and a single CF component is sufficient to represent each mode. For CFTD-SALT comparison, in \Fig{fig:2mode}(a) we plot the ``intensity'',  $\sum_n |\eCS_n^{(\mu)}|^2$ from SALT, and compare it to $I_m =  \frac{1}{T_2 - T_1} \int_{T_1}^{T_2} \text{d}t \, |\eC_m(t)|^2$, calculated for a sufficiently long sampling time after the steady state is reached in CFTD. 

The threshold of the first mode as calculated by SALT and our time-dynamical method is almost the same, and the emission intensities also coincide up to the point where a second mode begins to lase in the SALT calculation [see \Fig{fig:2mode}(a)].  Shortly thereafter, the second mode begins to lase in the time-dynamical calculation as well and both modes progress with comparable slope-efficiencies up to high pump powers. The steady-state frequencies [\Fig{fig:2mode}(b)] confirm the expected steady-state behavior. The small offset between the CFTD and SALT solutions can be attributed to the use of SVE in CFTD, whereas SALT does not make this approximation (See Ref.~\cite{ge_quantitative_2008} for a discussion of this point).


\begin{figure}[h]
\includegraphics[clip,width=1.0\linewidth]{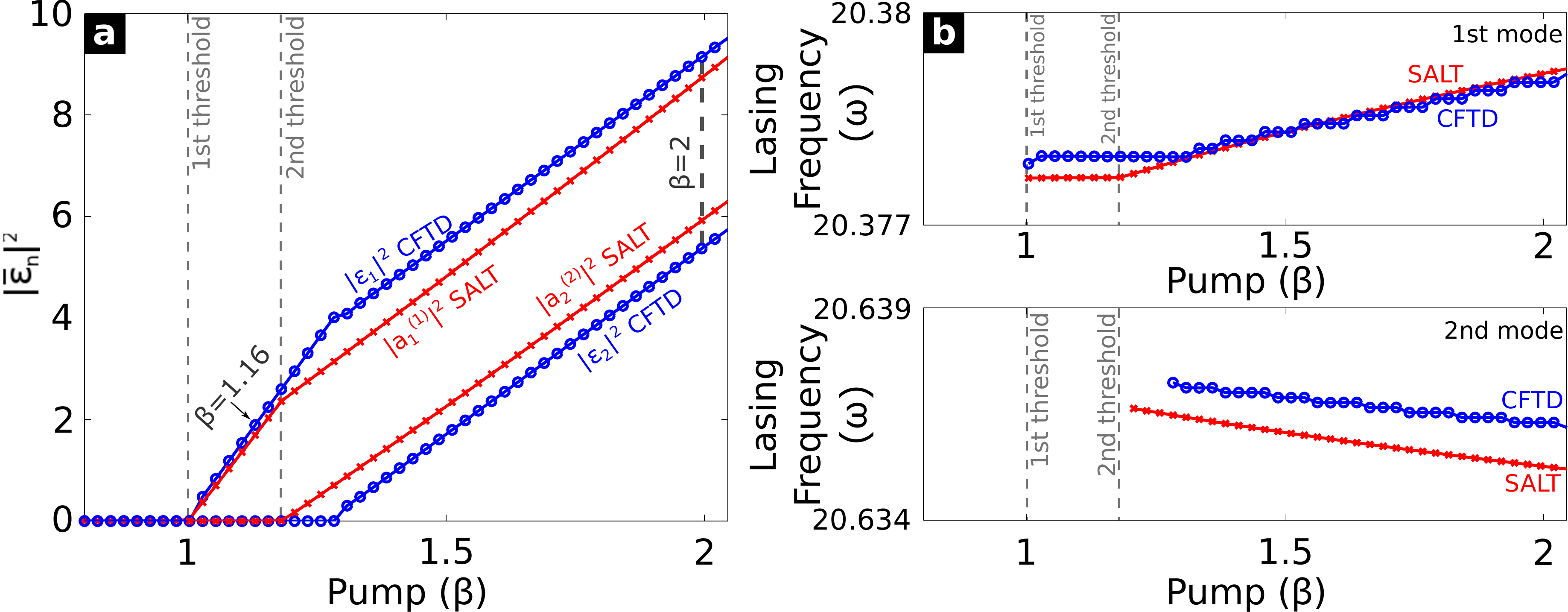}
\caption{(a) The steady-state emission intensity of the two lasing modes as calculated by SALT (red) and CFTD (blue). The values shown here are a time-average (details in text) of the steady-state time domain behavior at each pump step; the time-domain behavior is shown in \Fig{fig:CFTD}(a-b) for selected values of $\beta=1.16$ and $\beta=2$ (marked in this plot). (b) The steady-state center frequency for both lasing modes for SALT (red) and CFTD (blue). }
\label{fig:2mode}
\end{figure}

\subsection*{Regime B: Non-stationary inversion}

Keeping all other parameters, we now choose $\gpar=1$. At the onset of the second mode, this gives $\gpar/\Delta = 5.61$. We will find that $\Delta$ will change dramatically in this case, essentially going to zero as the pump power is increased. This phenomenon is known as  'cooperative frequency locking' \cite{lugiato_cooperative_1988}, and has been experimentally studied in Ref.~\cite{tamm_frequency_1988}. 

\begin{figure}[h]
\includegraphics[clip,width=1.0\linewidth]{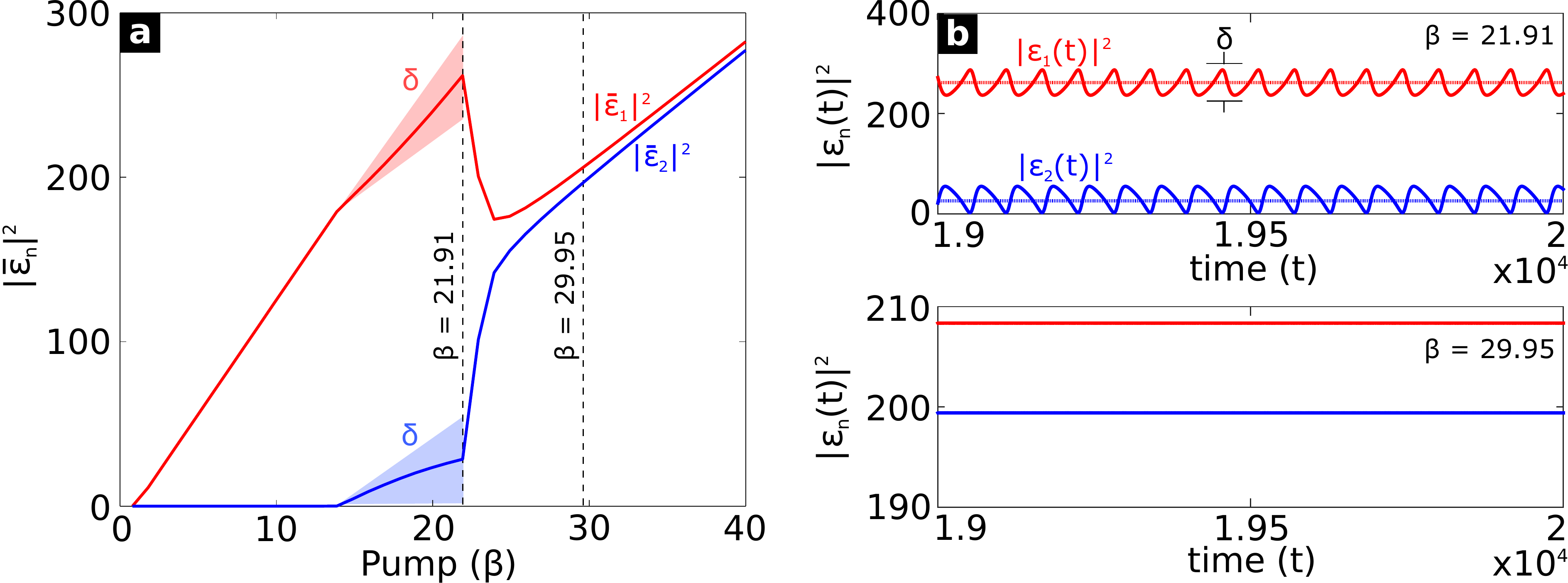}
\caption{(a) The steady-state emission intensity of the two lasing modes. The first mode (red) reaches threshold at the same pump power as \Fig{fig:2mode} while the second mode (blue) reaches threshold at more than ten times the threshold of the first mode. At $\beta = 21.91$ the two modes begin to converge, leading to a sharp rise (decline) in the blue (red) mode accompanied by oscillations steadily increasing in amplitude $\delta$ (shown by the light red and blue shaded triangular regions). After synchronizing, the modal intensities of the two modes are comparable and they increase linearly. (b) The top figure shows the large oscillations in time in each mode, immediately before synchronization; the straight lines identify the time-average of the oscillating signal and correspond to the values marked in (a). The bottom figure shows the absolute absence of these oscillations and a return to a steady-state after synchronization.}
\label{fig:sync1}
\end{figure}

In \Fig{fig:sync1}, CFTD reveals that while the first mode starts lasing at the same threshold as before, with a nominally identical lasing frequency $\Omega_1 = 20.3762$, the second mode lases at a threshold nearly 10 times larger with frequency $\Omega_2 = 20.5426$. However, immediately after the turn-on of the second mode, $|\varepsilon_n(t)|^2$ ceases to reach a stationary value, implying the existence of multiple frequencies in the respective spectra $\varepsilon_n(\omega)$. In lieu of intensities, we plot here the time-averaged quantities $I_m$, and indicate the size of oscillations, $\delta$, around the mean by shaded regions. 
As the pump approaches the synchronization threshold $D_{sync}^{th}$, the oscillations in the intensities grow (for the second mode, the oscillation magnitude remains always of the order of the mean, implying a clear limit cycle solution). For  $D_0 > D_{sync}^{th}$ the oscillations in the intensities abruptly disappear, and all field amplitudes oscillate at a single, synchronized frequency. The synchronization threshold is clearly defined and corresponds to $\beta = 22.98$. 

\begin{figure}[h]
\includegraphics[clip,width=1.0\linewidth]{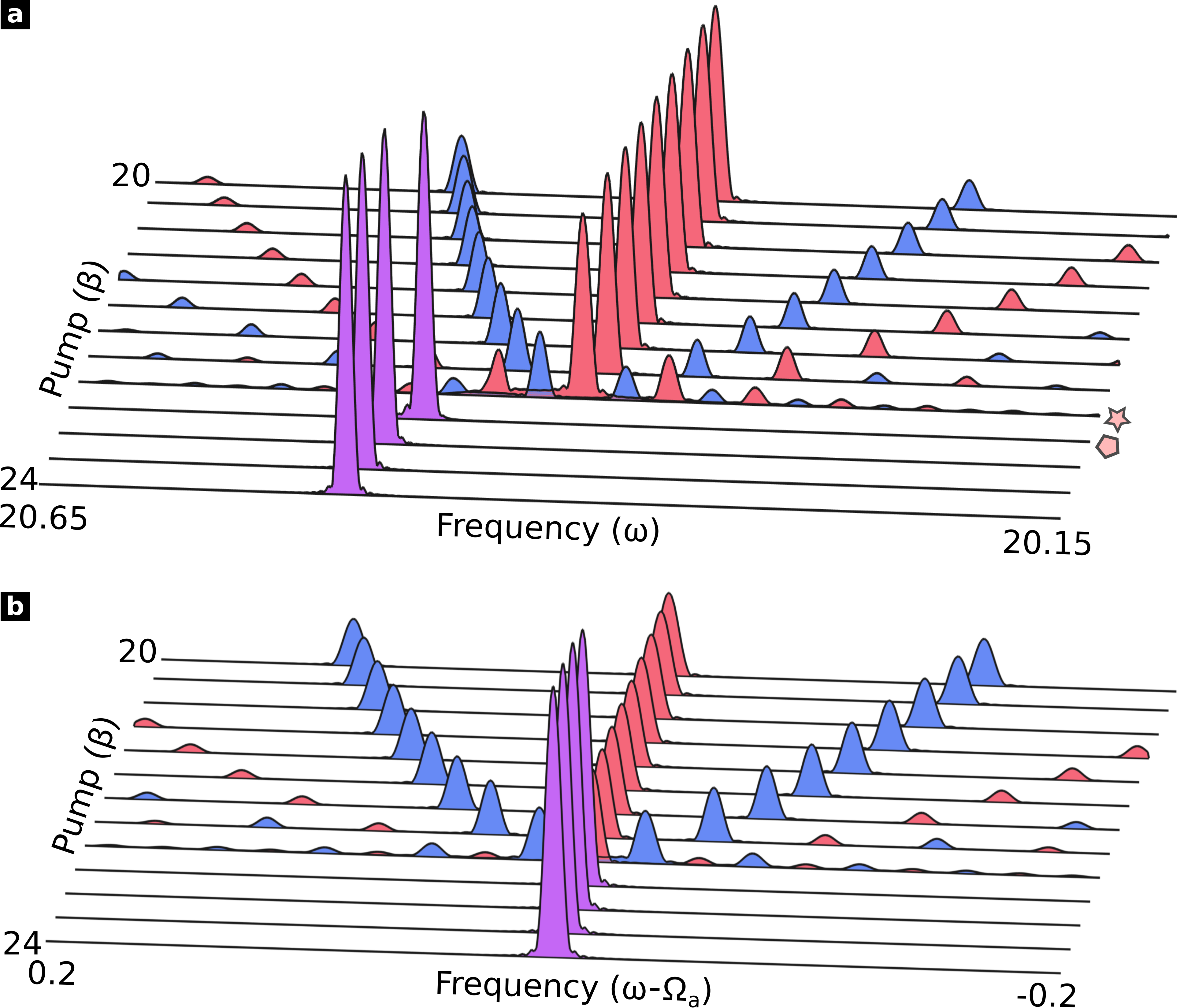}
\caption{(a) All red (blue) peaks belong to the spectrum of $\eC_1$ ($\eC_2$). All peaks are seen to draw closer and converge at $\beta = 22.64$ (marked by a star) and a single purple peak representing the nearly identical spectra of $\eC_1$ and $\eC_2$ can be seen at $\beta = 22.98 = D_{sync}^{th}$ (marked by a pentagon) and beyond. (b) The red (blue) peaks belong to the $D_{11}$ ($D_{12}$) component of inversion. Compared to the inset in \Fig{fig:CFTD}(d), the peaks of $D_{12}$ are comparable in magnitude to the DC peak from $D_{11}$, representing the significant effect of time-dynamical behavior in this calculation. Similar to (a), all peaks converge upon synchronization and only a DC component (purple) remains from $D_{11}$ and $D_{12}$.}
\label{fig:sync2}
\end{figure}

A close look at the dynamical behavior of the system near the synchronization threshold is provided in \Fig{fig:sync2}. We follow here the PSDs of (a) the electric field and (b) the $D_{11}$ and $D_{12}$ components of inversion as a function of the pump. Note that these spectra are shown for only a small range of pump powers around the synchronization threshold at $\beta = 22.98$. The largest peak in \Fig{fig:sync2}(a) belongs to the dominant mode, shown in red in \Fig{fig:sync1}(a), and the sub-dominant peak belongs to the mode shown in blue in \Fig{fig:sync1}(a). As pump power is increased, $\Delta$ decreases, and additional peaks enter the monitored frequency window, separated by integer multiples of $\Delta$ (with respect to the original laser frequencies $\Omega_i$). These are the frequency-doubling harmonics mentioned in \cite{tamm_frequency_1988}. Note that the highly non-linear sawtooth-like oscillations in the intensities seen in \Fig{fig:sync1}(b) are closely linked with this proliferation of frequencies in the power spectrum. As the pump power is increased further, all these peaks approach each other in a dramatic manner and at $\beta = 22.64$, recollect into the single peak shown (in purple) at $\beta = 22.98$ and beyond. This peak is seen to be shifted from the point of convergence and from both primary frequency components, and it is pulled towards the gain center. A slightly different perspective is offered by the evolution of the power spectrum of the inversion [\Fig{fig:sync2}(b)] which also shows that the off-diagonal frequency components (blue) converge into the DC component (red) as they must if there is to remain only one mode. The new mode that emerges beyond synchronization is comprised of nearly equal contributions from both CF states, which can be seen directly from the coming together of $|\eC_1|^2$ and $|\eC_2|^2$ in \Fig{fig:sync1}(a). The new mode has a non-trivial spatial pattern, which is embodied in a non-linearly generated phase between the two CF states composing the new laser mode. More detail on this point is provided in the Appendix.

\begin{figure}[h]
\includegraphics[clip,width=1.0\linewidth]{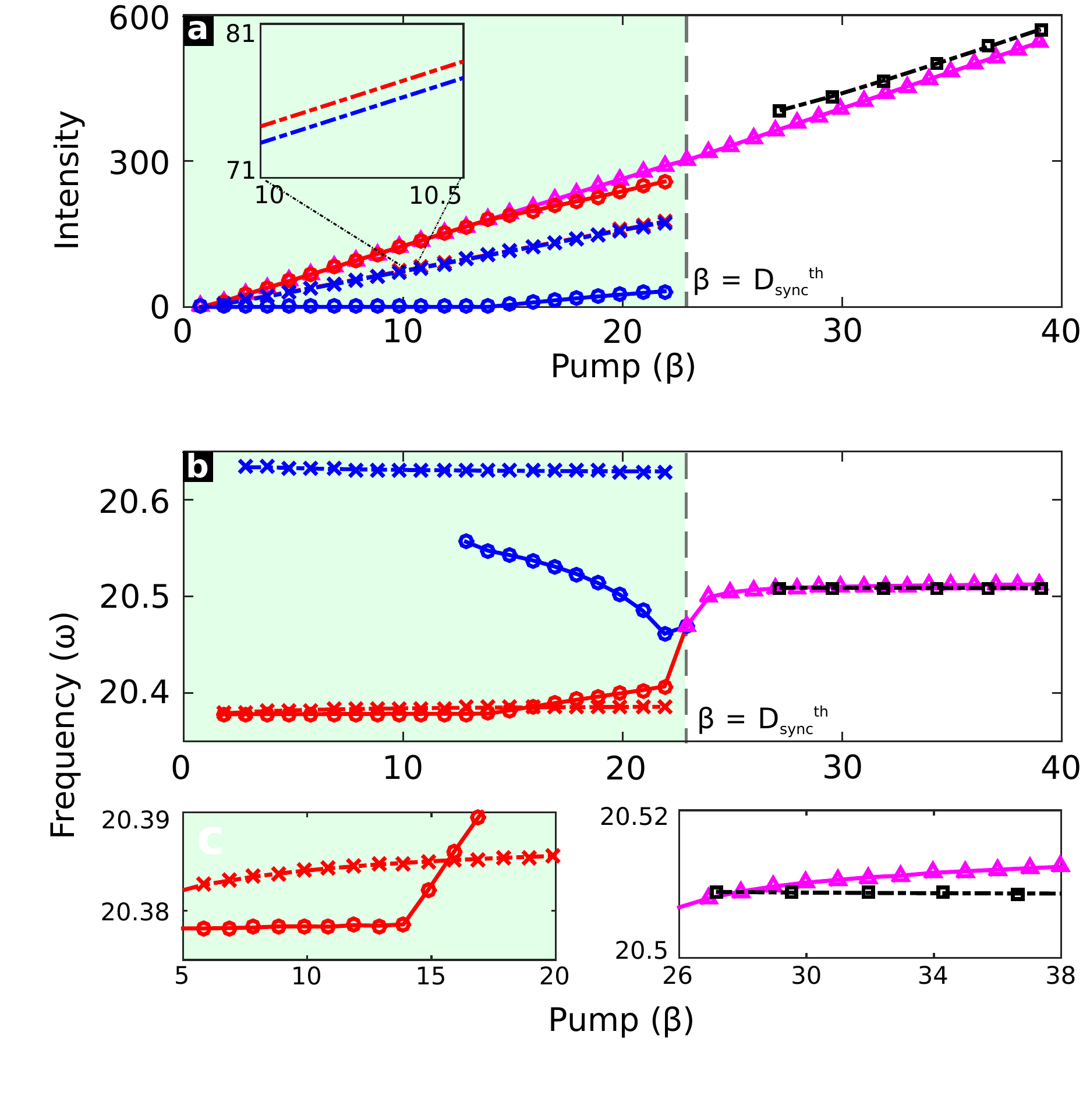}
\caption{SALT and CFTD calculations are compared for the same 2-mode calculation as above both before and after synchronization ($\gpar = 1$). (a) A comparison of the intensities as a function of pump. CFTD calculation for intensities of the two lasing modes is shown in red/blue circles while the SALT result is shown in red/blue crosses. Purple triangles represent the sum of the intensities of the two modes in the time-dependent calculation. The shaded green (white) region marks the pump range before (after) synchronization. Black squares show the SALT calculation after synchronization. (b) A comparison of the frequencies as a function of pump. Again, red/blue circles mark the time-dependent calculation and red/blue crosses mark the SALT calculation. Purple triangles (black squares) show the time-dependent (SALT) synchronized solution. The two smaller plots zoom in on regions of interest.}
\label{fig:SALT}
\end{figure}

It is interesting to see how all this looks from the perspective of SALT. We provide a comparative study in \Fig{fig:SALT}. The SIA appears to be valid everywhere outside the comparatively narrow range of pump powers $15 \lesssim \beta < 22.64$ and the SALT-CFTD correspondence should in principle be possible. Comparing the intensities in \Fig{fig:SALT} (a), however, we see what is a strinkingly large discrepancy between SALT and CFTD for $\beta \approx 15$. While SALT finds two modes that turn on relatively close to each other ($D_2^\text{th} \approx 1.3 D_1^\text{th}$), CFTD shows that the second mode does not turn on before $\beta = 12.86$. These seemingly disparate results should however be taken with a grain of salt. We first point out that the two thresholds found by SALT are identical to those found for $\gpar = 0.001$ shown in \Fig{fig:2mode}. This is of course expected because SALT equations do not depend on $\gpar$ when expressed in scaled variables (\Eq{eq:scalingfacts}), which is what is plotted in the vertical axis. The large discrepancy (despite SIA appearing to be valid) is simply because SALT predicts the turn-on of the second mode incorrectly, by a large margin. The consequence is that the change in slope of the intensity of the first mode that happens when the second mode turns on is incorrectly predicted by SALT as well. The seemingly large discrepancy between intensities by the time the second mode turns on in CFTD at $\beta = 12.86$ is thus simply due to the incorrect slope. We note that the pump range we are comparing is extremely large (and could well be inaccessibly large for certain gain media) -- the synchronization threshold found is about 22 times larger than the (lowest) laser threshold. 

The culprit for the incorrect prediction by SALT of the threshold of the second mode is interestingly still due to the breakdown of SIA, but in a non-trivial manner. While the oscillating corrections to the {\it inversion} are still small for $\beta \lesssim 15$, they generate a polarization component oscillating at $\Omega_1 + \Delta$ i.e. $\Omega_2$, that is proportional to the intensity of the first mode $\sim | \lasmodeS^{(1)}(\spc) |^2$ that does become large as pump power is increased. It can be shown that the threshold condition of the second mode is changed by a term proportional to $(\gpar/\Delta)  | \lasmodeS^{(1)}(\spc) |^2$. An appropriately modified set of two-mode SALT equations can be found \cite{ge_quantitative_2008}, and its implementation would correctly reproduce the behavior seen in CFTD for $\beta \lesssim 15$. 

However, SALT will have nothing to say and will fail {\it qualitatively} in capturing the physics in the range of pump powers plotted in \Fig{fig:sync2} very near the synchronization threshold. This is directly linked with the appearance of oscillating terms in the inversion (see \Fig{fig:sync2}(b)) that are comparable in magnitude to the static terms. Note that very interestingly the oscillations only appear in the off-diagonal elements, while diagonal elements mostly remain stationary. An analytic understanding of these features will be investigated in future work. 

Post-synchronization, as seen in \Fig{fig:SALT} for $\beta > D_{sync}^{th}$, a properly conditioned SALT (discussed in the Appendix) very precisely predicts the synchronized mode, both its oscillation frequency and the spatial composition. This again, is not surprising because now the SIA is valid to an excellent approximation in a single-frequency regime of lasing. We note however that SALT is unable to capture the synchronization threshold accurately.

\section{Conclusion}

Here we've presented a computationally and conceptually efficient approach to isolating and studying time-dependent effects in lasers. Using a spectral approach, we fully treat the open nature of lasers and integrate out the spatial variables, obtaining dynamical equations for the time-dependent coefficients describing the electric and polarization fields and the inversion. This delivers a highly scalable multi-mode framework for analyzing intrinsically non-stationary phenomena in open resonators of arbitrary spatial complexity, gain medium distribution, and pump profile. The simplest of such effects, mode synchronization, is studied here in a simple 1D cavity featuring pairs of closely spaced quasi-degenerate modes (small $\Delta$). With small enough $\gpar$, we obtain a stationary behavior once the transients die out, as postulated at the outset. At larger $\gpar$, non-stationary behavior is demonstrated in a narrow range of pump powers. We find that the stationary inversion approximation is largely valid in the parameter regimes investigated here, failing only in a narrow range of pump powers, for very large $\gpar/\Delta$, and for a special choice of the resonator structure. We expect that CFTD will find application in particular in modeling time-dependent phenomena in quasi-2D and 3D laser structures, because of its efficient spectral decomposition method that takes into account the openness of the underlying resonator structure in essentially an exact manner. 

\section{Acknowledgments}

This work is supported through NSF-ERC \# EEC-0540832 (MIRTHE). K.G.M. is supported by the People Programme (Marie Curie Actions) of the European Union's Seventh Framework Programme (FP7/2007-2013) under REA grant agreement number PIOF-GA-2011- 303228 (project NOLACOME).

\section{Appendix}

\begin{figure}[h]
\includegraphics[clip,width=1.0\linewidth]{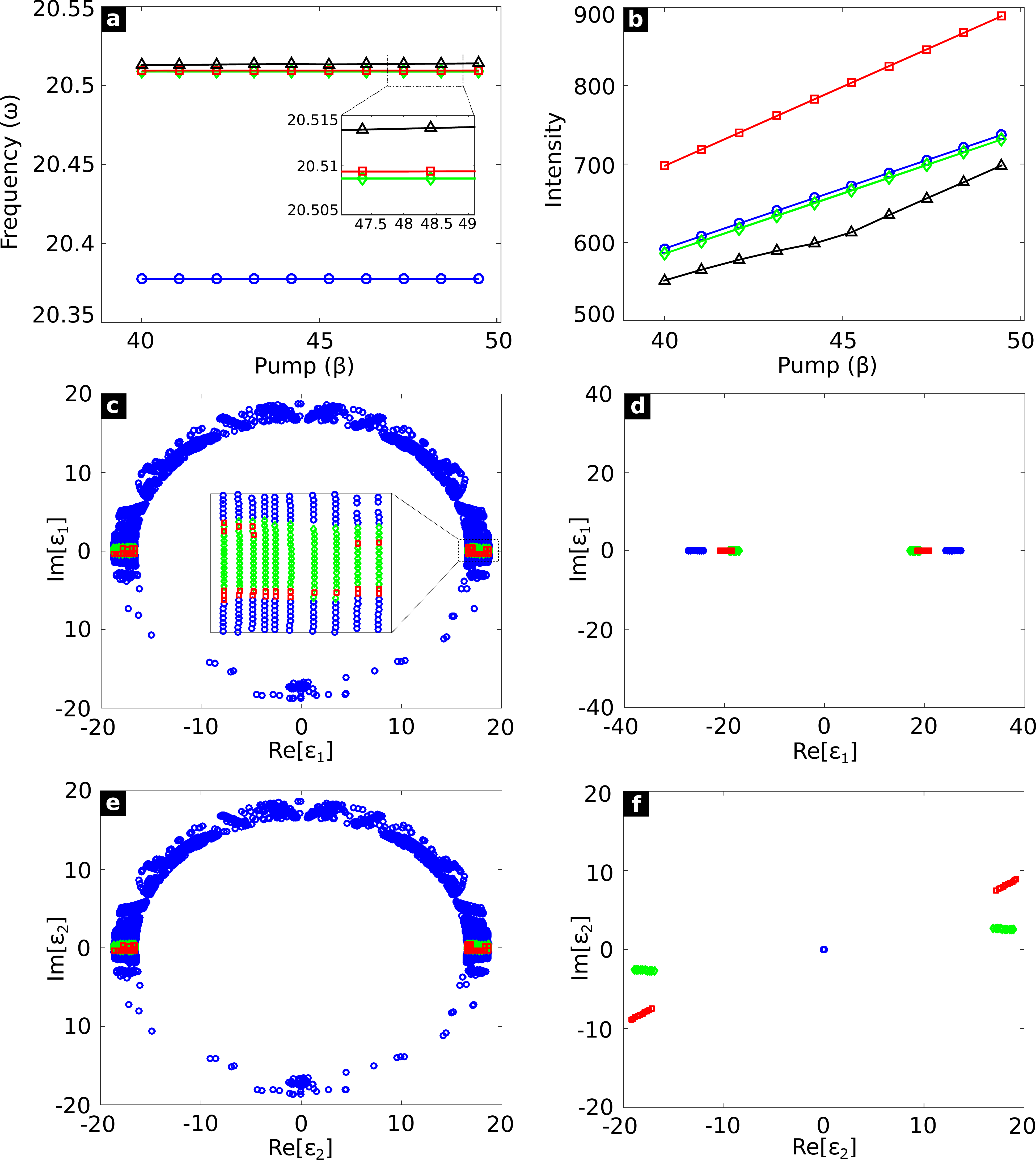}
\caption{(a) Frequencies as a function of the pump are shown for the seed (black triangles), and the three SALT solutions: the single-mode solution near the cavity frequency (blue circles), the synchronized solution (green diamonds), and the unstable solution (red squares). (b) Intensities are shown as a function of pump for the same color-scheme as (a). (c) and (e) The distribution of seeds in phase space is shown for $\eC_1$ and $\eC_2$; inset in (c) zooms in a region near the real axis where the synchronized and unstable solutions are found. (d) SALT solutions for $\eC_1$, which lie on the real line due to the SALT gauge condition. (f) SALT solutions for $\eC_2$ and their distribution in phase space.}
\label{fig:App}
\end{figure}

In this Appendix, our goal is to provide more detail on the SALT-CFTD correspondence in Regime B. The SIA is valid to a good approximation for $\beta \lesssim 15$ and $\beta > 22.98$, and a SALT-CFTD correspondence in these power ranges is therefore possible. 

As discussed in Regime B and \Fig{fig:SALT} above, in the pre-synchronization regime the apparent sizable discrepancy between SALT and CFTD solution is understood, and can be accounted for by a modified version of SALT \cite{ge_quantitative_2008}. We focus here on the SALT-CFTD correspondence in the synchronized regime, where it is important to properly condition SALT. 

The standard SALT algorithm uses an `adiabatic' sweep of pump power (not in the dynamical but computational sense). In other words, the solution in the previous step of pump power $D_0$ is fed as a seed for the non-linear solver for the next pump power. This practical procedure speeds up the computation in a dramatic way. If this were done blindly, then the two SALT solutions found in \Fig{fig:sync2} for $\beta<D_{sync}^{th}$ would simply extend without any apparent discontinuity to larger values of pump power, missing the synchronized solution. In fact, SALT has multiple single-frequency fixed points for $\beta>D_{sync}^{th}$, two of these are composed of a single CF mode (i.e. single pole) and the other two are composed of a particular balanced superposition of two CF modes (multi-pole solution). Interestingly, one of the latter is the synchronized solution found in the long-time limit of CFTD (shown using green markers in \Fig{fig:App}). This indicates that SALT does capture the fact that there is a single stable oscillation frequency and that the spatial structure of this laser mode is such that it is a particular superposition of the two spatial modes that were oscillating independently at lower powers. \Fig{fig:App}(d,f) indicate that three of these fixed points are stable, and one is unstable, as revealed with different initializations of the SALT non-linear solver (\Fig{fig:App}(c,e)). The unstable solution is a synchronized solution that is orthogonal to the stable synchronized solution. The stability of the two "single-pole" solutions (only one shown in the frequency window plotted) from the point of view of SALT is a perceived one, and is due to the neglect of the non-stationary terms in the inversion that in turn changes the stability structure of the solutions. We conclude that care must be exercised when conditioning SALT solutions for regimes outside its stated validity, even when the SIA appears to be a good approximation.



\bibliographystyle{unsrt}
\bibliography{hakan_bibtex}

%
%
%
%

\end{document}